\documentclass[twoside]{article}
\usepackage{fleqn}
\usepackage{espcrc2}

\usepackage{psfig}
\usepackage{graphicx}
\usepackage{psfrag}
\usepackage{amstex}

\newcommand{\UU}{{\bf U}}
\newcommand{\GG}{{\bf G}}
\newcommand{\SSS}{{\bf S}}
\newcommand{\TT}{{\bf T}}
\newcommand{\HH}{{\bf H}}
\newcommand{\mat}[4]{\left(%
\begin{array}{cc}#1&#2 \\#3&#4\end{array}\right)}

\title{Quantum Spin Formulation of the Principal 
Chiral Model}

\author{B. Schlittgen\address{Department of Mathematics,\\ 
        Massachusetts Institute of Technology (MIT), 
        Cambridge, MA 02139, U.S.A.}%
        and 
        U.-J. Wiese\address{Center for Theoretical Physics,
        Laboratory for Nuclear Science and Department of 
        Physics,\\
        Massachusetts Institute of Technology (MIT),
        Cambridge, MA 02139, U.S.A.}
        }

\begin{document}

\begin{abstract}
We formulate the two-dimensional principal chiral model 
as a quantum 
spin model, replacing the classical fields by quantum
operators acting in a Hilbert space, and introducing
an additional, Euclidean time dimension. Using coherent
state path integral techniques, we show that in the 
limit in which a large representation is chosen for the 
operators, the low energy excitations of the model 
describe a principal chiral model in three dimensions.
By dimensional reduction, the two-dimensional
principal chiral model of classical fields is recovered. 
\vspace{1pc}
\end{abstract}

\maketitle

\section{INTRODUCTION}
In \cite{ChWi97,BChW97} the construction of QCD as a 
quantum link model was proposed. In this paper we follow 
this new approach to 
quantising field theories, known
as {\em D-theory} to reformulate the principal chiral model.
In this approach, classical fields
are replaced by quantum operators.
In order to compensate for 
the loss in the number of degrees of freedom brought
about by
passing from continuous field variables to discrete
eigenspectra of the quantum operators, the theory is
formulated with an additional dimension, which later
disappears;
classical fields emerge as
low-energy excitations of the discrete variables, 
provided the $(d+1)$-dimensional theory has massless
excitations: When the extent of the extra dimension
becomes small in units of the correlation length
$\xi$, the $d$-dimensional
field theory emerges via dimensional reduction.
The guiding principles in formulating such a field theory 
are symmetry considerations. In particular, 
the quantum operators are Lie group generators
chosen so that the Hamiltonian has the same symmetries
on a quantum level as the original action has on a 
classical level.

The virtue of the {\em D}-theory formulation of 
field theories is that 
the quantum partition function
$Z=\mbox{Tr}\left( \exp\left(-\beta {\bf H}
\right)\right)$ is a trace over the {\em discrete} 
eigenvalues of 
the quantum operators 
that make up 
the Hamiltonian.
This promises for a much more efficient treatment by 
numerical techniques, and it is hoped that cluster 
algorithms similar to those used for Heisenberg model 
calculations can be developed.

In the remainder of this article, we formulate the 
principal chiral model as a quantum spin model.
Using coherent state path integral techniques
we find an effective action for the low-energy 
excitations of the model and finally we show how 
the two-dimensional principal chiral model of classical
fields emerges via dimensional reduction.

\section{PRINCIPAL CHIRAL MODEL, $d=2$}

\subsection{{\em D-Theory} Formulation}

The action of the two-dimensional continuum target theory
is the following:
\begin{equation} 
\!\!\!\!\!\!S[U]=\frac{1}{2g^2}\int\!\! d^2 x \mbox{Tr}
       \left(\partial_{\mu}U^{\dagger}(x)
       \partial_{\mu}U(x)\right), 
\end{equation}
where  the $U(x)$ are unitary $N\times N$ matrices.

On the lattice, derivatives are replaced by finite 
differences, and we are free to add and subtract 
constant terms in the action to obtain
a lattice regularised action of the form
\begin{equation}
S_{\text{lat}}[U]=
-\frac{1}{g^2}\sum_{<xy>} \mbox{Tr}\left(U_{x}^{\dagger}
U_{y}+V(U_{x}U_{x}^{\dagger})\right).
\end{equation}
Note that $U_{x}U_{x}^{\dagger}=I$, so that the 
potential term is only a constant. Here it has no 
influence on the physics, but it motivates a similar term 
in the {\em D}-theory Hamiltonian, which will then
have some effect on the low energy effective theory 
of the model.

The target theory has a
global $U(N)_{L}\otimes U(N)_{R}$ symmetry of the
form $U_x\to U_x'=LU_x R^{\dagger}$, where 
$L$ and $R$ are unitary matrices. It is known that
this symmetry breaks to a $U(N)_{V}$ symmetry
in the ground state.

Let us now replace the classical fields $U^{ij}_x$
by quantum operators $\UU^{ij}_x$ and write down a 
{\em D}-theory Hamiltonian, which evolves the 
two-dimensional system in an additional Euclidean 
time direction:
\begin{eqnarray}
\lefteqn{{\bf H}=\frac{J}{2N}\sum_{x,\mu}\left[\UU_{x}^{ij}
\left(\UU_{x+\mu}^{ij}\right)^{\dagger}+\UU_{x+\mu}^{ij}
\left(\UU_{x}^{ij}\right)^{\dagger}\right. } \hspace{1cm}
\notag \\
 & & \hfill \left.+V\left(\UU_{x}^{ij}\left(\UU_{x}^{kj}
\right)^{\dagger}\right)
+\mbox{h.c.}\right].
\end{eqnarray}
We would like this Hamiltonian to have an
$U(N)_{L}\otimes U(N)_{R}$ symmetry, i.e.\
$\left[ \GG^a_L,\HH\right]=
\left[ \GG^a_R,\HH \right] =0$,
where $\GG_L^a$ and  $\GG_R^b$ are  mutually commuting 
sets of $U(N)$ generators.
This can be realised by embedding 
$u(N)_L$ and $u(N)_R$ diagonally 
in the algebra of $U(2N)$ as 
follows: \cite{BChW97} 
Let $\{\lambda^a\}$ be matrices of the defining
representation of $su(N)$, with commutation
relations given by
$[\lambda^a,\lambda^b]=2if^{ab}_{c}\lambda^c$.
Then, 

\begin{gather}
\begin{align*}
\left[ \GG^a_L,\GG^b_L\right] =2if^{ab}_{c}\GG^c_L, 
& \quad	
\left[ \GG^a_R,\GG^b_R\right] =2if^{ab}_{c}\GG^c_R, 
\notag \\
\left[ \GG^a_R,\UU^{ij}\right] =\UU^{ik}\lambda^a_{kj}, 
& \quad
\left[ \GG^a_L,\UU^{ij}\right] =-\lambda^a_{ik}\UU^{kj},
\end{align*} \\
\begin{align}
\left[ \GG^a_R,\GG^b_L\right] &=
\left[ \TT,\GG^a_L\right]=\left[ \TT,\GG^a_R\right] =0, 
\notag \\
\left[ \TT,\UU^{ij}\right] &=2\UU^{ij}, 
\notag \\
\left[ \Re\UU^{ij},\Re\UU^{k\ell}\right]&=
\left[ \Im\UU^{ij},\Im\UU^{k \ell}\right] 
\notag \\
 &=-i\left( \delta_{ik}\Im \lambda^a_{j\ell}\GG^a_R
        +\delta_{j\ell}\Im\lambda^a_{ik}\GG^a_L\right), 
\notag \\ 
\left[ \Re \UU^{ij},\Im \UU^{k\ell}\right] &=
\notag \\
 &\hspace{-2cm}i\left( \delta_{ik}\Re\lambda^a_{j\ell}
\GG^a_R
-\delta_{j\ell}\Re\lambda^a_{ik}\GG^a_L+
\frac{2}{N}\delta_{ik}\delta_{j\ell} \TT \right).
\end{align}
\end{gather}

If we restrict ourselves to representations of $u(2N)$
which correspond to rectangular Young tableaux
with $N$ rows and $n_c$ columns, we can use 
a fermionic basis of rishons for our 
representation \cite{BChW97,ReSa89}:
\begin{align}
\SSS_i^j&=\GG^{ij}_L=
         \sum_{x} \left(\ell^{i\alpha\dagger}_{x}
\ell^{j\alpha}_{x}-\frac{n_c}{2}\delta^{ij}\right),   
\hspace{0.5cm} \notag \\
\SSS_{N+i}^{N+j}&=\GG^{ij}_{R}=
                 \sum_{x} \left(r^{i\alpha\dagger}_{x}
 r^{j\alpha}_{x}-\frac{n_c}{2}\delta^{ij}\right), 
\notag \\
\TT&=\sum_{x}\left( r^{i\alpha\dagger}_{x}r^{i\alpha}_{x}
      -\ell^{i\alpha\dagger}_{x}\ell^{i\alpha}_{x}\right), 
\notag \\
\SSS_{i}^{N+j}&=\UU^{ij}_{x}=
             \ell^{i\alpha}_{x}r^{j\alpha\dagger}_{x}, 
\notag \\
\SSS_{N+i}^{j}&=\left(\UU^{\dagger}_{x}\right)^{ij}=
\left(\UU_{x}^{ji}\right)^{\dagger}=r^{i\alpha}_{x}
\ell^{j\alpha\dagger}_{x}, \\
\sum_{i}\left(\ell^{i\alpha\dagger}\ell^{i\beta}\right. &+
              \left. r^{i\alpha\dagger}r^{i\beta}\right)=
\delta^{\alpha \beta}N, \label{eq:2} 
\end{align} 
where $\alpha=1,\ldots, n_c$ is an additional colour index 
and $i,j=1,\ldots,N$. For convenience, we have chosen
these generators not to be traceless. We then have
$\GG^{a}_{L(R)}=\lambda^{a}_{ij}\GG^{ij}_{L(R)}$.
The constraint (\ref{eq:2})
at each lattice point is needed to obtain the correct 
representation.  
For $J>0$, this model is antiferromagnetic and we consider
a bipartite lattice, made up of
sublattices {\em A} and {\em B}. For sites on
sublattice {\em A} we choose one representation, and 
for sites on sublattice {\em B} we choose its 
conjugate representation.
In our case, both representations share the same Young
tableau, so they are unitarily equivalent. But they 
need not be the same representation, as we shall see
in the next section.
{\em Note that the properties of $\,\HH$ are 
completely determined, once the 
representation of 
$u(2N)$ has been specified.} 
In particular, the physics will be independent of whether
the generators are represented by fermionic or bosonic
operators.

\subsection{Low Energy Effective Theory}

Following reference \cite{ReSa89} 
we can set up a coherent state path
integral of the form
$Z=\int {\cal D}Qe^{-S}$,
where $S=$
\begin{multline}
\!\!\!\frac{n_c}{4}\int_{0}^{\beta}\!\!\! d\tau\! \int_{0}^{1}
\!\!\! d u 
\left[\mbox{Tr}\left( Q(\tau,u)
\frac{\partial Q(\tau,u)}{\partial u}
\frac{\partial Q(\tau,u)}{\partial \tau}\right)\right] \\
 -\int_{0}^{\beta}d\tau H(Q(\tau)).
\end{multline}
Without going into the details, in this derivation
it is important to note that coherent states $|q\rangle$ 
are labeled by $GL(N,\mathbb{C})$ matrices $q$
and that $\langle q| \SSS_{\alpha}^{\beta}(x)|q\rangle
=\eta_x (n_c /2)Q_{\alpha}^{\beta}(x)$.  
[$\eta_x=+1(-1)$ for $x$ in sublattice $A$($B\,$).]
The Q-field is of the form:
\begin{multline} \label{eq:1}
Q= \exp\left[\mat{0}{q}{-q^{\dagger}}{0}\right] \\
\times \mat{I_N}{0}{0}{-I_N}
\exp\left[\mat{0}{-q}{q^{\dagger}}{0}\right],
\end{multline}
where $q\in GL(N,\mathbb{C})$. The following boundary
conditions hold:
\begin{gather}
Q(\tau,0)=Q(\tau',0), \quad \text{for all}
                               \;\tau,\tau'; \notag \\
Q(\tau,1)=Q(\tau);  \quad
Q(0,u)=Q(\beta ,u).
\end{gather}

We now decompose the matrix field $q$ into a unitary 
matrix field $U$ and a hermitian matrix field $C$:
$q=CU$.
Under $U(N)_L \otimes U(N)_R$ transformations, these
fields transform as follows:
$q\to q'=LqR^{\dagger}$, 
$U\to U'=LUR^{\dagger}$ 
and $C\to C'=LCL^{\dagger}$.
Substituting this coset decomposition into (\ref{eq:1})
we find,
\begin{equation}
Q=\mat{\cos(2C)}{-\sin(2C)U}{-U^{\dagger}\sin(2C)}
{-U^{\dagger}\cos(2C)U}.
\end{equation}

Now define: $B\equiv -\sin (2C)$ and
choose $V[(-\sin(2C))UU^{\dagger}(-\sin(2C))]=V(B^2)$ 
such that 
its coefficients are of order one and the minimum
of $-3B^2+V(B^2)$ is attained for $B$ of the form
$B=bI,\; 0<b<1$.
While all other terms in the Hamiltonian are proportional
to $a^2$ ($a$ is the lattice spacing), the term 
$-3B^2+V(B^2)$ is not, so that when taking the continuum
limit $a\to 0$, any fluctuations of this term around
its minimum value are suppressed. 
The field $B$ is thus frozen
in the value $bI$.

Next, we decompose the field $U(x)$ into staggered and
uniform components:
\begin{gather}
U(x)\approx \Omega(x)\sqrt{1-a^2 L^{\dagger}(x)L(x)}
+\eta_x aL(x), \notag \\
\Omega(x)L(x)+L^{\dagger}(x)\Omega(x)=0,
\end{gather}
and $\Omega(x)$ is unitary. Then integrate out the
$L$ field to find
an effective action for the 
long-wavelength uniform fluctuations of the form
\begin{multline}
S=\frac{n_c}{2}\eta_{x}\sum_{x}\int_{0}^{\beta}
d\tau \mbox{Tr}\left( \sqrt{1-b^2}\Omega \partial_{\tau}
\Omega^{\dagger}\right) \\ 
+\int_{0}^{\beta}\!\!\!d\tau \int \!\! d^2 x
\frac{\rho_s}{2}\mbox{Tr}\left( \partial_{\mu}\Omega
\partial_{\mu}\Omega^{\dagger} +
\frac{1}{c^2}\partial_{\tau}\Omega
\partial_{\tau}\Omega^{\dagger}\right).
\end{multline}
Here, $\rho_s=Jb^2n_c^2/2N$ is the spin stiffness and 
$c=Jb^2 n_c a/(N\sqrt{1-b^2})$ is the spin wave velocity.
The first term is a Berry Phase term.

We therefore get a low energy effective action for the 
Goldstone modes associated with the symmetry breaking 
pattern $U(N)_L\otimes U(N)_R\to U(N)$.

Finally, we assume that the correlation length 
is much larger than the extent of the additional
time dimension, $\xi\gg \beta c$, 
so that the system is dimensionally reduced to
a two-dimensional model. We can then integrate over
$\tau$.  From \cite{Wieg84}, we have that for the
two-dimensional system, 
$\xi\propto \exp(2\pi/(g^2 N))=
\exp(2\pi\beta\rho_s/N)$,
which is indeed consistent with $\xi\gg \beta c$
in the zero-temperature $(\beta\to\infty)$ limit.

\section{SUMMARY AND CONCLUSIONS}

We have reformulated the principal chiral model as a 
quantum spin model, in which the fields take on discrete 
values (the eigenvalues of the operators). To compensate
for this restriction, we introduced an extra time direction.
Having chosen a particular representation for the 
operators in the Hamiltonian, we were then able to show 
that 
the dimensionally reduced theory emerges from the low-energy
theory of the collective excitations of the discrete 
variables.
We note that the representation chosen for the
operators influences the symmetry breaking pattern.

The extension of these results to
quantum link QCD is currently under investigation.

The virtue of the D-theory formulation of field theories 
is that 
numerical simulations should be much easier than in 
conventional field theory formulations, due to the discrete
nature of the field variables.


\end{document}